%% file: main.tex
\documentclass[sigconf,screen]{acmart}

\copyrightyear{2021}
\acmYear{2021}

\acmConference[SAC'21]{the 36th ACM/SIGAPP Symposium on Applied Computing}{March, 2021}{Gwangju, Korea}
\acmArticle{4}
\acmPrice{15.00}
\acmISBN{978-1-4503-XXXX-X/18/06}

\usepackage{amsmath}
\usepackage[toc,page]{appendix}
\usepackage{comment}
\usepackage{multirow}
\usepackage{graphicx}
\graphicspath{{./Figs/}}
\usepackage{subfigure}
\usepackage{url}

\usepackage{listings}
\usepackage{xcolor}
\usepackage{verbatim}
\usepackage{wasysym}
\usepackage{textcomp}

\usepackage{amssymb}
\usepackage{pifont}
\usepackage{ulem}
\usepackage{appendix}
\usepackage{anyfontsize}
\usepackage{subfigure}
\usepackage{booktabs}
\usepackage{longtable}

\usepackage{array}
\usepackage{rotating}
\newcommand{\PreserveBackslash}[1]{\let\temp=\\#1\let\\=\temp}
\newcolumntype{C}[1]{>{\PreserveBackslash\centering}p{#1}}
\newcolumntype{R}[1]{>{\PreserveBackslash\raggedleft}p{#1}}
\newcolumntype{L}[1]{>{\PreserveBackslash\raggedright}p{#1}}
\usepackage[linesnumbered,lined, noend, algoruled,norelsize]{algorithm2e}
\usepackage{mathtools}
\usepackage{flushend}
\usepackage{algorithm2e}
\usepackage{seqsplit}
\usepackage{indentfirst}
\SetAlFnt{\small}
\SetAlCapFnt{\small}
\SetAlCapNameFnt{\small}

\begin{document}
\title{CLUE: Towards Discovering Locked Cryptocurrencies in Ethereum}

\author{Xiaoqi Li}
\affiliation{%
	\institution{Department of Computing, The Hong Kong Polytechnic University}
	\city{Hong Kong SAR}
	\country{China}}
	\email{csxqli@gmail.com}
	
\author{Ting Chen}
\affiliation{%
	\institution{Center for Cybersecurity, University of Electronic Science and Technology of China}
	\city{Chengdu}
	\country{China}}
\email{brokendragon@uestc.edu.cn}

\author{Xiapu Luo}
\authornote{The corresponding author.}
\affiliation{%
	\institution{Department of Computing, The Hong Kong Polytechnic University}
	\city{Hong Kong SAR}
	\country{China}}
\email{csxluo@comp.polyu.edu.hk}

\author{Chenxu Wang}
\affiliation{%
	\institution{School of Software Engineering, Xi’an Jiaotong University}
	\city{Xi’an}
	\country{China}}
\email{cxwang@mail.xjtu.edu.cn}



\input{./Sections/abs}

\ccsdesc[500]{Security and privacy~Distributed systems security}
\ccsdesc[500]{Software and its engineering~Software usability}
\ccsdesc[500]{Security and privacy~Software security engineering}

\keywords{Distributed System Security, Ethereum, Cryptocurrency}


\maketitle

\input{./Sections/intro}

\input{./Sections/background}

\input{./Sections/system}
\input{./Sections/evaluation}

\input{./Sections/discussion}

\bibliographystyle{IEEEtran}
\normalem
\bibliography{ref}

\end{document}

%% file: Sections/abs.tex
\begin{abstract}
\label{sec:abs}
As the most popular blockchain that supports smart contracts, there are already more than 296 thousand kinds of cryptocurrencies built on Ethereum. However, not all cryptocurrencies can be controlled by users. For example, some money is permanently locked in wallets' accounts due to attacks. In this paper, we conduct the \textit{first} systematic investigation on locked cryptocurrencies in Ethereum. In particular, we define three categories of accounts with locked cryptocurrencies and develop a novel tool named \textsc{Clue} to discover them. Results show that there are more than 216 million dollars value of cryptocurrencies locked in Ethereum. We also analyze the reasons (i.e., attacks/behaviors) why cryptocurrencies are locked. Because the locked cryptocurrencies can never be controlled by users, avoid interacting with the accounts discovered by \textsc{Clue} and repeating the same mistakes again can help users to save money.
\end{abstract}

%% file: Sections/intro.tex
\section{Introduction}
\label{sec:intro}
As the most popular blockchain that supports smart contracts, there are many kinds of contract-based cryptocurrencies built in Ethereum. Apart from ETH, which is the native cryptocurrency of Ethereum, more than 296 thousand cryptocurrency contracts are deployed in Ethereum~\cite{0}. These cryptocurrencies have high market capitalization. For example, the ETH has a total value of about 20 billion dollars~\cite{1}, and USDT has a total value of more than four billion dollars~\cite{2}. Note that all the cryptocurrencies' prices in this paper are based on statistics in September, 2020~\cite{0}.

However, not all cryptocurrencies can be controlled by users. Actually, much value of cryptocurrency is permanently locked in some accounts. For example, the attacker escalated his privilege and destructed Parity's multi-sig library contract in 2017~\cite{li2020glaser}, which locked all the ETH stored in Parity wallet accounts. Through our analysis, there are 203 wallet accounts with more than 515,035 locked ETH, which is worth more than 192 million dollars. Many users still sent cryptocurrencies to the attacked wallet accounts, leading more money permanently lost. If the accounts with locked cryptocurrencies can be detected and alerted in time, users can reduce their economic losses.

Unfortunately, there still lacks systematic research on the locked cryptocurrencies in Ethereum. To fill this gap, we propose and develop a novel tool named \textsc{Clue} (disCovering Locked cryptocUrrency in Ethereum), which can discover three categories of accounts with more than 216 million dollars value of locked cryptocurrencies. In particular, we discover two categories of contract accounts with locked cryptocurrencies due to contract destruction or attacks, and one category of EOAs (Externally Owned Accounts) with locked cryptocurrencies due to users' unreasonable behaviors. Note that calling to accounts with locked cryptocurrencies not only wastes system computation resources, but also wastes users' money.

The main contributions of this paper are as follows:

(1) To the best of our knowledge, we conduct the \textit{first} research that systematically analyzes locked cryptocurrencies in Ethereum. We propose and define three categories of accounts with locked cryptocurrencies, i.e., one kind of EOAs and two kinds of smart contract accounts. 

(2) We implement a tool named \textsc{Clue} to detect each category of accounts with locked cryptocurrencies. For smart contract accounts, we analyze their account states in StateDB and analyze their historical transactions, to discover destructed contracts. Leveraging symbolic execution, we analyze the runtime bytecodes of smart contracts to discover attacked Parity wallet contracts. For EOAs, we mainly use account state analysis and transaction analysis to detect contract-creation failure EOAs.

(3) We analyze the attacks/behaviors related to the discovered locked cryptocurrencies, which can explain why they are locked and help users to save money. We also conduct experiments to evaluate its quantity and accuracy. A total of 216,186,551.12\textdollar~value of cryptocurrencies are discovered by \textsc{Clue}, and all of the discovered cryptocurrencies are permanently locked in Ethereum.

%% file: Sections/background.tex
\section{Background and Related Work}
\label{sec:back}

\textbf{Ethereum:} Ethereum is the most popular blockchain system that supports smart contracts~\cite{5}. There are two kinds of accounts in Ethereum, i.e., EOA and contract account~\cite{li2020stan}. EOA is controlled by user through its private key, which does not store any code. Contract account is created by EOA or another contract, which stores the runtime bytecodes of the contract. Smart contract is a program deployed and executed in blockchain~\cite{chen2018understanding}. Every node in Ethereum runs an EVM (Ethereum Virtual Machine) and the runtime bytecodes are executed in the EVM. When a user calls the smart contract, he/she needs to send transaction with gas to the address of the target contract~\cite{chen2018towards}. Every operation of contract runtime bytecodes consumes specific amount of gas when they are executed in the EVM~\cite{chen2017under}. Developers/users can also destruct the deployed smart contract through executing \texttt{SELFDESTRUCT} operation~\cite{chen2020}.

\textbf{Cryptocurrency:} 
\label{sec:cryp}
Cryptocurrency is digital assets based on blockchain techniques~\cite{6}. There are two categories of cryptocurrencies in Ethereum, i.e., ETH and CBC (Contract-Based Cryptocurrency)~\cite{6}. ETH is the native cryptocurrency of Ethereum. Apart from ETH, there are many other kinds of cryptocurrencies based on contracts, and ERC20 is the most popular standard of CBC~\cite{li2020stan}. All the CBC analyzed in this paper are compliant with ERC20. Both EOA and contract account can hold cryptocurrency. EOA can transfer out ETH by initiating transactions from it, and contract can transfer out ETH by executing specific operations (e.g., \texttt{CALL}, \texttt{SELFDESTRUCT}). Note that accounts can only transfer out their CBC by calling the corresponding ERC20-based smart contract. The ERC20 standard provides some basic functions and events that must be implemented of CBC in Ethereum. If the user $U_a$ wants to transfer out CBC, $U_a$ can call \texttt{transfer()}. Furthermore, the user $U_a$ can authorize another account $U_b$ to transfer out CBC through calling \texttt{transferFrom()}. Before $U_b$ transfers out $U_a$'s CBC, $U_a$ must authorize the account $U_b$ through calling \texttt{approve()}.

\textbf{Account State:} Every account in Ethereum has four state fields stored in StateDB (State DataBase)~\cite{li2020glaser}. For each account $a$, we mainly analyze three fields. Code $\sigma{[a]}_c$ stores the smart contract's runtime bytecodes, which is empty if $a$ is an EOA. Balance $\sigma{[a]}_b$ stores the ETH balance value (in Wei) of the account. Nonce $\sigma{[a]}_n$ stores the number of transactions \textit{sent from} EOA, or the number of contracts created by contract account.

\textbf{Related Work:} Chen et al.~\cite{17} detected Ponzi schemes, which are classic frauds and might cheat users' ETH. They built a classification model to detect latent Ponzi schemes by using data mining and machine learning methods. Cheng et al.~\cite{18} analyzed the attack that steals cryptocurrencies exploiting unprotected JSON-RPC endpoints. They designed and implemented a honeypot that could capture real attacks in the wild. Ji et al.~\cite{ji2020deposafe} implemented a tool named \textsc{Deposafe} to detect and exploit the fake deposit vulnerability in ERC-20 tokens. However, all of the above work analyzed the cryptocurrencies illegally possessed by criminals, and they did not analyze locked cryptocurrencies that does not belong to anyone. \cite{19} measured the network properties and structures of ERC20 smart contracts, and \cite{6} analyzed inconsistent behaviors in ERC20 smart contracts. However, they focused on analyzing the smart contracts' implementations and invocations, whose purposes differ from ours. There are some other work analyzed cryptocurrencies in Ethereum~\cite{frowis2019detecting} or other blockchains~\cite{howell2020initial}.

%% file: Sections/system.tex
\begin{figure*}[ht]
	\centering
	\vspace*{-2ex}
	\includegraphics[width=6.00in]{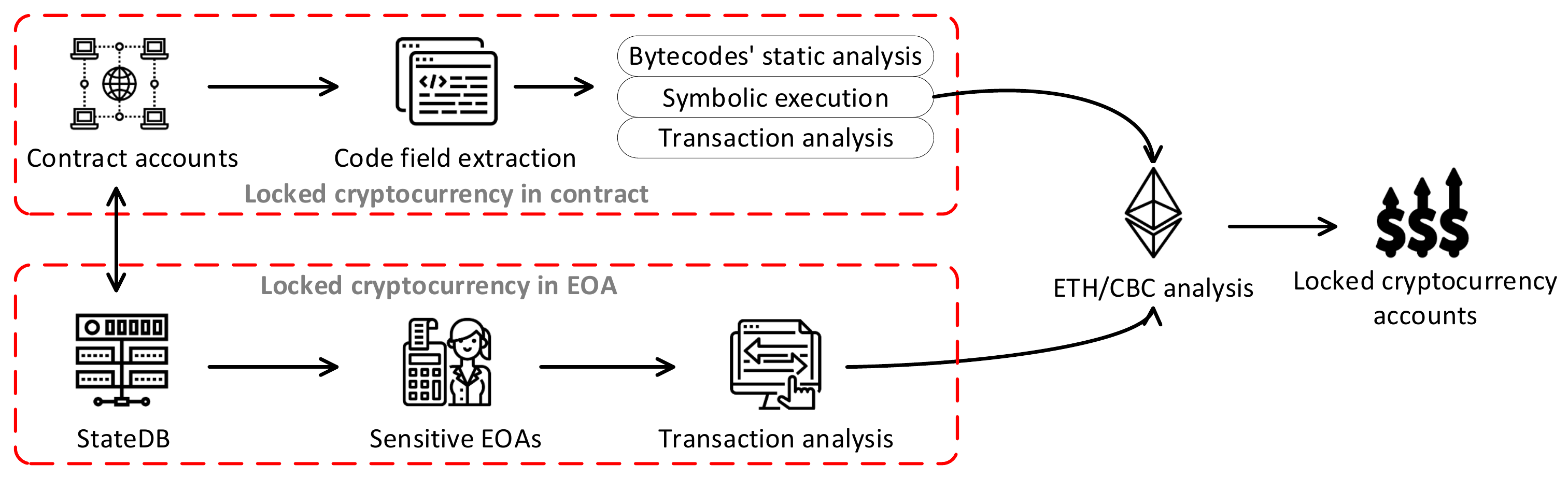}
	\vspace*{-1ex}
	\caption{Overview of \textsc{Clue}'s architecture. There are mainly two modules in \textsc{Clue}: detection of locked cryptocurrencies in contract and detection of locked cryptocurrencies in EOA.}
	\vspace{-2ex}
	\label{architecture}
\end{figure*}

\section{\textsc{Locked Cryptocurrencies}}
\label{sec:lockedc}

\textbf{Destructed Contract:} In Ethereum, the smart contract can be destructed and transfer out all its stored ETH through executing the \texttt{SELFDESTRUCT} operation. After destruction, the smart contract account will be deleted from StateDB. However, some users may not know in time of the smart contract's destruction and still send ETH/CBC to it, which leads to the sent ETH/CBC be locked. After sending ETH to the destructed smart contract account, the contract account with the same address before destruction will be created again in the StateDB. For the CBC held by the destructed contract account, most of it will also be permanently locked. Because the destructed contract account stores no runtime bytecodes, it cannot send out transaction. Therefore, the destructed contract account cannot transfer out its CBC through calling \texttt{transfer()} function, or authorize another account to transfer out its CBC through calling the \texttt{transferFrom()} function. Above all, all the ETH and most of the CBC held by the destructed contract accounts are permanently locked in Ethereum. For example, one smart contract named \textsc{Insightsnetworkcontributions} (Address: \href{https://etherscan.io/address/0x97eC9BFb0F6672C358620615a1E4dE0348Aea05c}{\seqsplit{0x97eC9BFb...}}) is discovered by \textsc{Clue} as destructed contract with locked cryptocurrencies. It has been transferred more than 208 ETH after its destruction, which is worth more than 79 thousand dollars.

\textbf{Attacked Parity Contract:} In 2017, the attacker escalated his privilege and destructed the multi-sig library of Parity wallets, leading to all the ETH and most of the CBC held by wallet contracts that depend on the library locked permanently~\cite{li2020glaser}. The attacker destructed the wallets' library in the following process. First, the attacker called the library's functions \texttt{initWallet()} and \texttt{initMultiowned()} through the fallback function, to escalate his/her privilege. Second, the attacker destructed the library contract through calling function \texttt{kill()}. After the library's destruction, all the wallet contracts can no longer call the library and executing its functions. Therefore, all the ETH stored in the attacked Parity wallet contracts is permanently locked. Furthermore, all the CBC held by the attacked wallet contracts is also locked. This is because the wallet contract cannot call the ERC20 contracts. For example, one attacked Parity wallet contract (Address: \href{https://etherscan.io/address/0x0da3cB3046F72fcbb49edF01B04AB6efc6C0D8DC}{\seqsplit{0x0da3cB30...}}) discovered by \textsc{Clue} stores 2,576.35ETH. After the attack, there were 17.88ETH transferred to this wallet contract. Furthermore, there is also 2.09\textdollar~value of CBC locked in it. If these wallet accounts with locked cryptocurrencies can be detected and alerted in time, the users might no longer transfer cryptocurrencies to it, which can help users to save money.

\textbf{Contract-creation Failure EOA:} 
\label{sec:creationcryp}
When the user deploys a smart contract in Ethereum, he/she will still receive one fake contract address if the contract-creation fails. Indeed, the received contract address does not exist in StateDB just after the contract-creation failure. However, some users might wrongly ignore the failure message and still transfer cryptocurrencies to the fake contract address, leading to cryptocurrencies locked permanently. Because the address with locked cryptocurrencies never stores code, we classify it as EOA. For Example, one EOA (Address: \href{https://etherscan.io/address/0x5488b0a000843dc54b0e541dfb75c2927f92adc8}{\seqsplit{0x5488b0a0...}}) discovered by \textsc{Clue} locks 19 ETH in value of \texttildelow7,088.71 dollars and some CBC in value about seven dollars. After the user encountered an \textit{out-of-gas} error during contract-creation, he still called the fake contract address three times.

\section{\textsc{Clue}}
\label{sec:clue}
The overview of \textsc{Clue}'s architecture is shown in Figure~\ref{architecture}, which mainly consists of two modules: \underline{(1)} Locked cryptocurrencies in contracts. For destructed contract, we debug accounts' historical transactions and detect destructed contracts through transaction trace analysis and balance analysis. For attacked Parity contract, we statically analyze contracts' runtime bytecodes and detect wallet contracts through symbolic execution. \underline{(2)} Locked cryptocurrencies in EOAs. We export sensitive EOAs from StateDB and detect contract-creation failure accounts with locked cryptocurrencies through transaction and balance analysis.

\textbf{Detection of Destructed Contracts:} 
\label{clue:sc}
The detection of destructed contracts with locked cryptocurrencies is divided into four steps. First, for accounts stored in the StateDB, we debug their historical external transactions through Geth API \texttt{debug.traceTransaction()}. From the execution trace of the external transaction, we analyze whether it ever executed the \texttt{SELFDESTRUCT} operation, which is used for destructing the contract account. Second, for the external transaction that executed \texttt{SELFDESTRUCT}, we leverage Ethereum RPC API to get the detailed information of the transaction. Because the execution of \texttt{SELFDESTRUCT} will produce internal transaction, we get the detailed information of the internal transaction according to the hash of external transaction. Third, leveraging the transaction's execution trace and detailed information, we analyze the specific address of the destructed contract account. If the type field of one internal transaction is ``\texttt{suicide}'', we can conclude that it is used for destructing the contract account. Then we export the sender address of the internal transaction, which is the address of destructed contract. Fourth, we analyze ETH/CBC balance of the destructed contract through Ethereum RPC-APIs. At last, the destructed contracts with locked cryptocurrencies can be discovered.

\textbf{Detection of Attacked Parity Contracts:} 
\label{clue:sc-parity}
The detection of attacked Parity contracts with locked cryptocurrencies is divided into four steps. First, for contract accounts stored in the StateDB, we export their runtime bytecodes from the code field $\sigma{[a]}_c$. Second, we statically analyze the bytecodes. In particular, we use \textsc{Disasm}~\cite{10} to disassemble the runtime bytecodes and detect hardcoded Parity library pattern. Third, we leverage symbolic execution techniques to analyze the runtime bytecodes with the hardcode pattern. We use \textsc{Oyente}~\cite{luu2016making} as the symbolic execution engine. During the symbolic execution process of runtime bytecodes, we monitor the external call related operations. If we encounter external call operation's execution, we analyze its second operand $P_a$, which is used for the target address of the external call. If $P_a$ is a real value and equals with the hardcoded Parity library's address, we can conclude that the corresponding analyzed contract account is an attacked Parity contract. Furthermore, the attacked Parity contract cannot call ERC20 contracts to transfer out its CBC. This is because $P_a$ does not equal with ERC20 contracts' addresses or associated with transaction's input data. Fourth, we analyze the ETH/CBC balances for the detected contracts in the third step.  

\textbf{Detection of Contract-creation Failure EOAs:} 
\label{clue:eoa}
We leverage account state analysis and transaction analysis to detect contract-creation failure EOAs with locked cryptocurrencies, which is divided into three steps. First, we traverse the StateDB and filter out sensitive EOAs. The sensitive EOAs have the following state features: nonce $\sigma{[a]}_n$ is zero, and code $\sigma{[a]}_c$ is empty. The sensitive EOAs with these features never send out any transaction. $\sigma{[a]}_c$ field is empty indicates that the account $a$ is an EOA. For an EOA, its $\sigma{[a]}_n$ field stores the number of transactions sent from it. Second, leveraging Ethereum RPC-API, we fetch and analyze sensitive EOA's oldest transaction, to verify that it encountered an error and returned a smart contract address. As described in Section~\ref{sec:creationcryp}, the contract-creation transaction will also return a fake contract address when it fails with errors. Third, we analyze ETH/CBC balance of the detected EOAs in the second step through Ethereum RPC-APIs. At last, contract-creation failure accounts with locked cryptocurrencies can be discovered.

%% file: Sections/evaluation.tex
\section{Evaluation}
\label{sec:eva}
We carry out experiments to answer the following research questions: \underline{RQ1 (Quantity):} How much value of locked cryptocurrencies can be detected by \textsc{Clue}? \underline{RQ2 (Accuracy):} To what extent can \textsc{Clue} accurately discover locked cryptocurrencies?

\begin{table}[ht!]
	\centering
	\vspace*{-2ex}
	\scriptsize
	\caption{Statistics of locked cryptocurrencies and related accounts detected through \textsc{Clue}. (\ding{109}: discovered candidate accounts. \ding{108}: accounts with locked cryptocurrencies.)}
	\vspace{-3ex}
	\label{tab_quantity}
	\begin{tabular}{|c|c|c|c|}
		\hline
		\textbf{Category}&\textbf{Discovered account}&\textbf{Locked ETH}&\textbf{Locked CBC}\\ \hline
		
		Destructed contract&5,916,076\ding{109} | 173\ding{108}&123,841.02\textdollar&25,036,305.09\textdollar \\ \hline
		
		Attacked Parity contract&658\ding{109} | 203\ding{108}&190,060,328.19\textdollar&950,380.79\textdollar \\ \hline
		
		Contract-creation failure EOA&3,720\ding{109} | 191\ding{108}&15,640.76\textdollar&55.27\textdollar \\ \hline
		
		\textit{Total}&5,920,454\ding{109} | 567\ding{108}&190,199,809.97\textdollar&25,986,741.15\textdollar \\ \hline
	\end{tabular}
	\vspace{-2ex}
\end{table}
\textbf{RQ1 Quantity:} We evaluate the quantity of locked cryptocurrencies detected by \textsc{Clue}, whose statistics are shown in Table~\ref{tab_quantity}. Applying \textsc{Clue} to all Ethereum StateDB data, we totally discover 216,186,551.12\textdollar~value of locked cryptocurrencies. The related accounts' addresses for each category and analyzed transaction data are published on\url{https://figshare.com/articles/dataset/11605296}. For the destructed contracts, many of them were created due to DoS attacks~\cite{chen2017adaptive}. The attacker created large amount of smart contracts and destructed them through \texttt{SELFDESTRUCT} operation. Most of these destructed contracts are not called any more by normal users. Therefore, most of the destructed contracts do not lock any cryptocurrency. For the attacked Parity contracts, we totally discover 658 related accounts, while Etherscan only tags 153 of them. For contract-creation failure EOAs, their locked cryptocurrencies' value is small, because users might stop calling these accounts after they realize the contract-creation failure. The locked CBC of destructed contracts does not be transferred out during contracts' destruction, which leads to more locked CBC than ETH. Furthermore, all these detected accounts might lock more cryptocurrencies with Ethereum's running, and we also plan to measure locked cryptocurrencies' time accumulation in our future work. \textbf{Answer to RQ1:} For the proposed three kinds of Ethereum accounts, we totally discover \underline{216,186,551.12\textdollar} value of cryptocurrencies locked in them.

\textbf{RQ2 Accuracy:} For destructed contracts, we check all the 173 discovered accounts through Etherscan. All of them have been tagged \textsl{``Self-Destruct''}, and they all have more than zero value of ETH/CBC. Furthermore, there is no ETH/CBC transferred out after their destruction. Similarly, all the 203 attacked Parity wallets have more than zero value of ETH/CBC, and their ETH/CBC never be transferred out after the Parity attack (Transaction hash: \href{https://etherscan.io/tx/0x47f7cff7a5e671884629c93b368cb18f58a993f4b19c2a53a8662e3f1482f690}{\seqsplit{0x47f7cff7...}}). In addition, we decompile these contracts leveraging \textsc{Panoramix}~\cite{16}, and they all call the attacked Parity wallets' library. For the contract-creation failure accounts, we check all the 191 discovered accounts that lock cryptocurrencies through Etherscan. All of these accounts encountered errors during contract-creation, and they all have more than zero value of ETH/CBC. Also, their ETH/CBC is never transferred out. \textbf{Answer to RQ2:} 100\% of the 567 accounts discovered by \textsc{Clue} store cryptocurrencies, and all of these cryptocurrencies are locked permanently.

%% file: Sections/discussion.tex
\section{Discussion and Conclusion}
\label{sec:conc}
\textbf{Discussion:} We propose and detect three categories of accounts with locked cryptocurrencies in Ethereum, while there might also exist other categories. In our future work, we plan to analyze more categories of accounts with locked cryptocurrencies. We run \textsc{Clue} on all the Ethereum accounts' data. Although the number of discovered accounts with locked cryptocurrencies is small (i.e., 567), the value of locked cryptocurrencies is great. To the best of our knowledge, there is still no research of how many accounts with locked cryptocurrencies exist in Ethereum, and our work fills this gap. In other blockchain systems (e.g., Bitcoin, EOS), there might also exist locked cryptocurrencies. We plan to detect more cryptocurrencies in other blockchain systems in future work.

\textbf{Conclusion:} In this paper, we analyzed cryptocurrencies locked permanently in Ethereum. We defined three categories of accounts with locked cryptocurrencies and implemented a tool named \textsc{Clue}, which discovered more than 216 million dollars value of locked cryptocurrencies. We also analyzed why these cryptocurrencies are locked, which can help users/developers to avoid losing money.

%% file: main.bbl
\begin{thebibliography}{10}
\providecommand{\url}[1]{#1}
\csname url@samestyle\endcsname
\providecommand{\newblock}{\relax}
\providecommand{\bibinfo}[2]{#2}
\providecommand{\BIBentrySTDinterwordspacing}{\spaceskip=0pt\relax}
\providecommand{\BIBentryALTinterwordstretchfactor}{4}
\providecommand{\BIBentryALTinterwordspacing}{\spaceskip=\fontdimen2\font plus
\BIBentryALTinterwordstretchfactor\fontdimen3\font minus
  \fontdimen4\font\relax}
\providecommand{\BIBforeignlanguage}[2]{{%
\expandafter\ifx\csname l@#1\endcsname\relax
\typeout{** WARNING: IEEEtran.bst: No hyphenation pattern has been}%
\typeout{** loaded for the language `#1'. Using the pattern for}%
\typeout{** the default language instead.}%
\else
\language=\csname l@#1\endcsname
\fi
#2}}
\providecommand{\BIBdecl}{\relax}
\BIBdecl

\bibitem{0}
``Token tracker,'' \url{https://etherscan.io/tokens}, 2020.

\bibitem{1}
``Total ether supply and market capitalization,''
  \url{https://etherscan.io/stat/supply}, 2020.

\bibitem{2}
``Tether usd,''
  \url{https://etherscan.io/token/0xdac17f958d2ee523a2206206994597c13d831ec7},
  2020.

\bibitem{li2020glaser}
X.~Li, T.~Chen, X.~Luo, and J.~Yu, ``Characterizing erasable accounts in
  ethereum,'' in \emph{Proc. of ISC}, 2020.

\bibitem{5}
X.~Li, P.~Jiang, T.~Chen, X.~Luo, and Q.~Wen, ``A survey on the security of
  blockchain systems,'' in \emph{Future Generation Computer Systems}, 2020.

\bibitem{li2020stan}
X.~Li, T.~Chen, X.~Luo, T.~Zhang, L.~Yu, and Z.~Xu, ``Stan: Towards describing
  bytecodes of smart contract,'' in \emph{Proc. of QRS}, 2020.

\bibitem{chen2018understanding}
T.~Chen, Y.~Zhu, Z.~Li, J.~Chen, X.~Li, X.~Luo, X.~Lin, and X.~Zhange,
  ``Understanding ethereum via graph analysis,'' in \emph{Proc. of INFOCOM},
  2018.

\bibitem{chen2018towards}
T.~Chen, Z.~Li, H.~Zhou, J.~Chen, X.~Luo, X.~Li, and X.~Zhang, ``Towards saving
  money in using smart contracts,'' in \emph{Proc. of ICSE}, 2018.

\bibitem{chen2017under}
T.~Chen, X.~Li, X.~Luo, and X.~Zhang, ``Under-optimized smart contracts devour
  your money,'' in \emph{Proc. of SANER}, 2017.

\bibitem{chen2020}
T.~Chen, Y.~Feng, Z.~Li, H.~Zhou, X.~Luo, X.~Li, X.~Xiao, J.~Chen, and
  X.~Zhang, ``Gaschecker: Scalable analysis for discovering gas-inefficient
  smart contracts,'' in \emph{IEEE Transactions on Emerging Topics in
  Computing}, 2020.

\bibitem{6}
T.~Chen, Y.~Zhang, Z.~Li, X.~Luo, T.~Wang, R.~Cao, X.~Xiao, and X.~Zhang,
  ``Tokenscope: Automatically detecting inconsistent behaviors of
  cryptocurrency tokens in ethereum,'' in \emph{Proc. of CCS}, 2019.

\bibitem{17}
W.~Chen, Z.~Zheng, J.~Cui, E.~Ngai, P.~Zheng, and Y.~Zhou, ``Detecting ponzi
  schemes on ethereum: Towards healthier blockchain technology,'' in
  \emph{Proc. of WWW}, 2018.

\bibitem{18}
Z.~Cheng, X.~Hou, R.~Li, Y.~Zhou, X.~Luo, J.~Li, and K.~Ren, ``Towards a first
  step to understand the cryptocurrency stealing attack on ethereum,'' in
  \emph{Proc. of RAID}, 2019.

\bibitem{ji2020deposafe}
R.~Ji, N.~He, L.~Wu, H.~Wang, G.~Bai, and Y.~Guo, ``Deposafe: Demystifying the
  fake deposit vulnerability in ethereum smart contracts,'' in \emph{arXiv
  preprint}, 2020.

\bibitem{19}
S.~Somin, G.~Gordon, and Y.~Altshuler, ``Network analysis of erc20 tokens
  trading on ethereum blockchain,'' in \emph{Proc. of ICCS}, 2018.

\bibitem{frowis2019detecting}
M.~Fr{\"o}wis, A.~Fuchs, and R.~B{\"o}hme, ``Detecting token systems on
  ethereum,'' in \emph{Proc. of FC}, 2019.

\bibitem{howell2020initial}
S.~T. Howell, M.~Niessner, and D.~Yermack, ``Initial coin offerings: Financing
  growth with cryptocurrency token sales,'' in \emph{The Review of Financial
  Studies}, 2020.

\bibitem{10}
``Disasm,'' \url{https://github.com/Arachnid/evmdis}, 2019.

\bibitem{luu2016making}
L.~Luu, D.-H. Chu, H.~Olickel, P.~Saxena, and A.~Hobor, ``Making smart
  contracts smarter,'' in \emph{Proc. of CCS}, 2016.

\bibitem{chen2017adaptive}
T.~Chen, X.~Li, Y.~Wang, J.~Chen, Z.~Li, X.~Luo, M.~H. Au, and X.~Zhang, ``An
  adaptive gas cost mechanism for ethereum to defend against under-priced dos
  attacks,'' in \emph{Proc. of ISPEC}, 2017.

\bibitem{16}
``Panoramix,'' \url{https://github.com/eveem-org/panoramix}, 2019.

\end{thebibliography}
